\documentclass[onecolumn,preprintnumbers,amsmath,amssymb,elsart]{revtex4}

\usepackage{mathrsfs}

\usepackage{graphicx}
\usepackage{dcolumn}
\usepackage{bm}
\usepackage[center]{subfigure}

\begin{document}


\title{Nonclassical lattice solitons in optical lattice via Electromagnetically induced transparency}

\author {Yongyao Li$^{1,2}$}
\email{autumnlyy@gmail.com}
\author{Zhonghui Yuan$^{1}$}
\author{Wei Pang$^{3}$}
\email{kingprotoss@gmail.com}
\author{Yikun Liu$^{2}$}

\affiliation{$^{1}$Department of Applied Physics, South China
Agricultural University, Guangzhou 510642, China\\
$^{2}$State Key Laboratory of Optoelectronic Materials
and Technologies,\\Sun Yat-sen University, Guangzhou 510275, China\\
$^{3}$Department of Experiment, Guangdong University of Technology,
Guangzhou 510006, China}


\begin{abstract}
An optical four-level atomic discrete system through optical
induction is proposed. A theoretical scheme to produce nonclassical
lattice solitons (NLS) in the system is presented with the use of
the effects of enhanced self-phase modulation and the giant kerr
effect in the electromagnetically induced transparency. The power
density and the photon flux can be tuned to a very low level by the
controlling field and the soliton can propagate with very slow group
velocity. By changing the sign of the detuning $\Delta_{1}$, both
in-phase and $\pi$ out-of-phase NLSs can be produced in this system.
\end{abstract}

\maketitle

Optical fields propagating in couple waveguide arrays exhibit novel
phenomena \cite{Christodoulides2003}. The light field shows great
functionality in such an optical periodic system which are
impossible appeared in the bulk. Recently, the behavior of
nonclassical light field in couple waveguide arrays becomes an
interesting issue \cite{Stefano,Armit}. Nonclassical light is always
a light field with single or few photons. In quantum information and
quantum communication, photons are widely used as the information
carrier to transfer signals. Generally, the number of photons are
very few \cite{Sherson}. Therefore, such a optical periodic system
provides an avenue to inventing new optical devices for quantum
information processing. To avoid signals degradation during
propagation, it is highly desirable to propagate the optical wave as
a kind of soliton. Discrete or lattice soliton (LS)
\cite{Lederer,Efremidis} is a soliton mode in such a periodic
waveguide arrays. It shows great potential in photonic network
communications \cite{Ablowitz}. However, in the nonclassical light
limit, the intensity of the soliton must be very low (i.e. at least
few photons per soliton cross section per nanosecond). Traditional
discrete systems such as kerr nonlinearity media \cite{yaron1998},
liquid crystals \cite{Fratalocchi} photorefractive crystals
\cite{Fleischer2} are very difficult to produce such kind of soliton
because a required large nonlinear refractive index is always
associated with intense light field. Therefore, it is important to
realize a new optical periodic system that can produce a good
controllability NLS in guiding these signals.

In this work, we will apply the technique of electromagnetically
induce transparency (EIT) to overcome this difficulty in the atomic
system. EIT media, which can be applied to inducing dissipation-free
strong photon-photon interactions, can offer far larger third-order
Kerr susceptibility and far smaller linear susceptibility than
conventional nonlinear media \cite{Fleischhauer2005}. Recently,
ultraweak \cite{THong} and ultraslow \cite{LDeng2} spatial and
temporal solitons are realized by utilizing EIT in uniform atoms
system, respectively. This opens the possibility of archiving NLS in
atomic discrete systems.

In EIT medium, a dissipation-free optical discrete system can be
generated by using optical induction technique. The energy diagram
of the scheme is shown in FIG. 1 (a). It is similar to the
traditional N-type system: the parties of $|1\rangle$, $|2\rangle$
are same and different from $|3\rangle$, $|4\rangle$. A weak probe
wave $E_{P}$ with Rabi-frequency $\Omega_{P}=\wp_{31}E_{P}/\hbar$ is
acting on a resonant transition $|1\rangle\rightarrow|3\rangle$ with
a single photon detuning $\Delta_{1}$. A running-wave field with a
Rabi-frequency $\Omega_{C}$ is driving the atomic transition
$|2\rangle\rightarrow|3\rangle$ with a detuning
$\Delta_{C}=\Delta_{1}$, in other works, the two-photon detuning
$\delta=\Delta_{1}-\Delta_{C}=0$. An optical induction field with a
Rabi-frequency $\Omega_{S}$ is inducing the transition
$|2\rangle\rightarrow|4\rangle$ with a detuning $\Delta_{2}$.
According to the previous studies \cite{Fleischhauer2005}, when the
probe exactly at the two-photon resonance ($\delta=0$) and the atoms
are all in the ground state, the absorption to the probe is vanished
and independent to the single-photon detuning.  Meanwhile, the
existence of the detuning $\Delta_{1}$ and $\Delta_{2}$ makes the
probe experience an enhanced SPM produced by itself \cite{XiaoM} and
XPM by $\Omega_{S}$ \cite{Schmidt}, respectively.

\begin{figure} 
\centering \subfigure[]{
\label{fig_1_a} 
\includegraphics[scale=0.18]{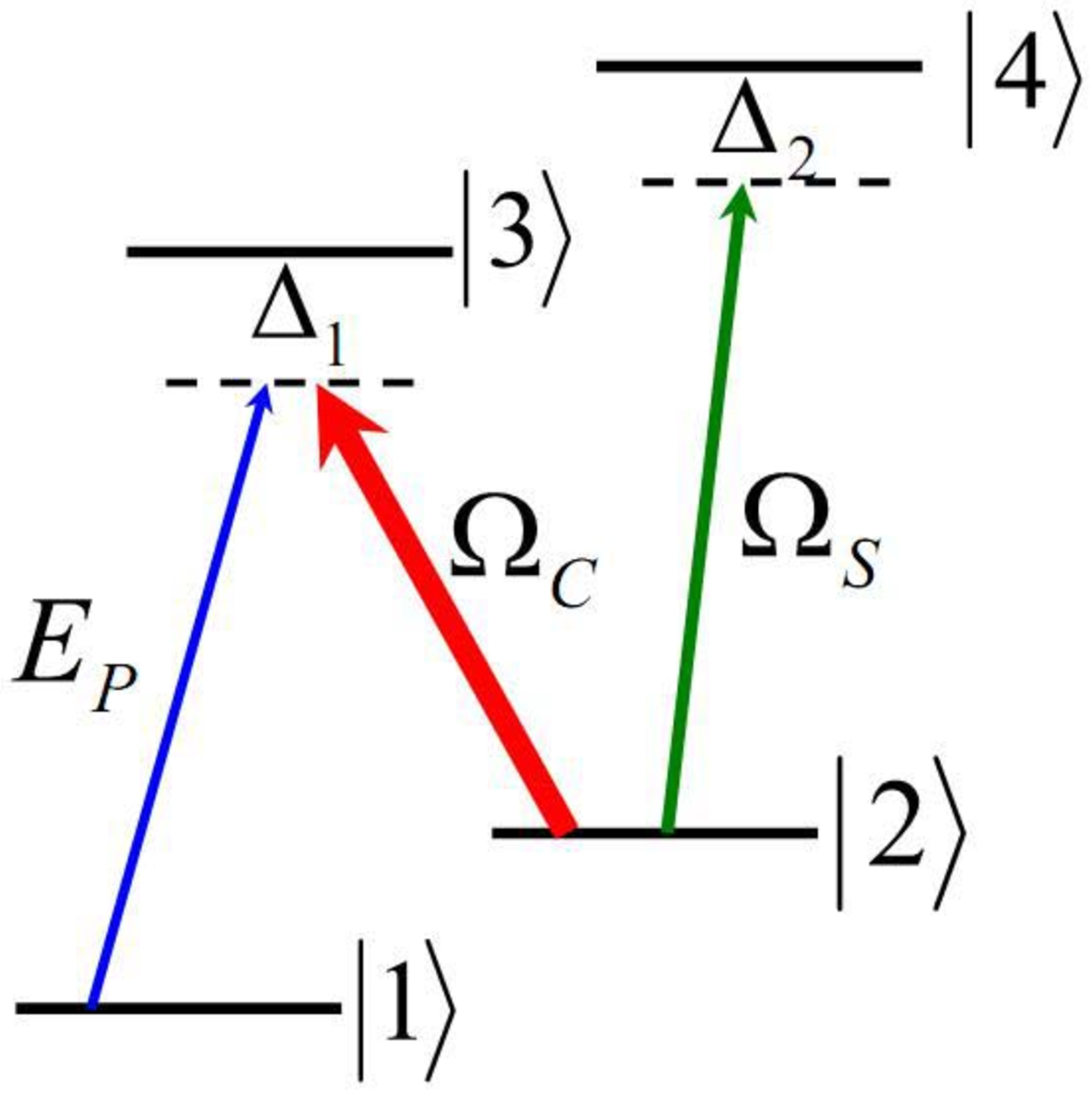}}
\hspace{0.02in} \subfigure[]{
\label{fig_2_b} 
\includegraphics[scale=0.45]{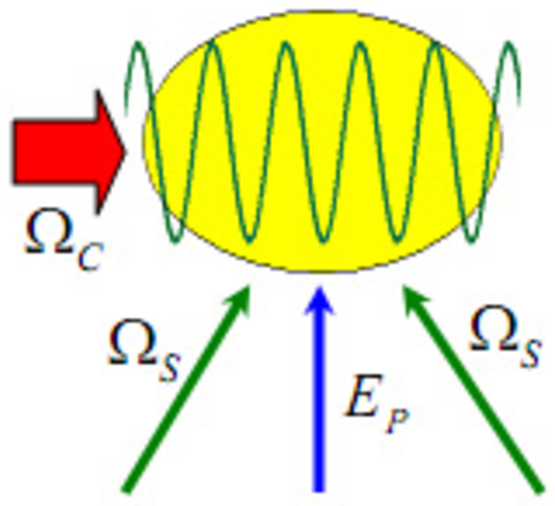}}
\caption{\label{fig:wide} (a) Energy diagram of the system. (b)
Geometric configuration of the lights. Lattice-forming beams
$\Omega_{S}$ and nonclassical light beam $\hat{E}_{P}$ are
co-propagating, The coupling light $\Omega_{C}$ is a p-wave,which is
propagating transversely along the atoms} \label{fig_1}
\end{figure}

We assume that $\Omega_{P},\Omega_{S}\ll\Omega_{C}$, which results
in the action of $\Omega_{P}$ and $\Omega_{S}$ can be treat as a
perturbation. The interaction Hamiltonian between the light and
atoms can be write as: $\hat{H}=\hat{H}_{0}+\hat{H}_{1}$, where
\begin{eqnarray}
\hat{H}_{0}=\sum^{5}_{i=1}\hbar\omega_{i}|i\rangle\langle
i|-{1\over2}[\Omega_{C}e^{-i\omega_{C}t}|3\rangle\langle2|+H.c],
\end{eqnarray}
is the Hamiltonian operator without perturbation and
\begin{eqnarray}
\hat{H}_{1}=-{1\over2}[\Omega_{P}e^{-i\omega_{p}t}|3\rangle\langle1|+\Omega_{S}|4\rangle\langle2|+H.c].
\end{eqnarray}
is the perturbation's Hamiltonian. Therefore, the evolution equation
of density matrix reads:
\begin{eqnarray}
\dot{\rho}^{(n)}=-{i\over\hbar}[\hat{H}_{0},\rho^{(n)}]-{i\over\hbar}[\hat{H}_{1},
\rho^{(n-1)}]-{1\over2}[\Gamma, \rho^{(n)}]_{+},
\end{eqnarray}

Where $n$ designates the $n$-th step of the density matrix. The
matrix elements $\langle i|\Gamma|j\rangle=\gamma_{i}\delta_{ij}$,
where $\gamma_{i}$ is the decay rate designating the population
damping from the energy level $|i\rangle$. Assuming that all the
atoms are populating at the ground state, in other words,
$\rho^{(0)}_{11}=1$ and $\rho^{(0)}_{ij}=0$ ($i$ or $j\neq1$), and
then the steady state solution of any step of the density matrix can
be deduced by Eq. (3). Here we only deduce to the 3-step, the higher
steps of the density matrix are ignored for their influences are
tiny to the system. In the deduction, we assume that
$\gamma_{2}\approx\gamma_{1}\approx 0$ and
$\gamma_{3}\approx\gamma_{4}=\gamma$. Furthermore, to avoid the
absorption from the atoms, we let $\Delta_{1}, \Delta_{2}\gg\gamma,
\Omega_{C}$. From Eq. (1)-(3) and the initial condition of
$\rho^{(0)}$, one can obtain that:
\begin{eqnarray}
\rho^{(1)}=\left(
\begin{array}{cccc}
 0 & \rho^{(1)}_{12} & 0 & 0 \\
 \rho^{(1)}_{21} & 0 & 0 & 0 \\
 0 & 0 & 0 & 0 \\
 0 & 0 & 0 & 0
\end{array}
\right)
\end{eqnarray}
and:
\begin{eqnarray}
\rho^{(2)}=\left(
\begin{array}{cccc}
 0 & 0 & 0 & \rho^{(2)}_{14} \\
 0 & 0 & \rho^{(2)}_{23} & 0 \\
 0 & \rho^{(2)}_{32} & 0 & 0 \\
 \rho^{(2)}_{41} & 0 & 0 & 0
\end{array}
\right)
\end{eqnarray}
where
$\rho^{(1)}_{21}=\rho^{(1)\ast}_{12}\approx-\Omega_{P}\Omega^{\ast}_{C}/|\Omega_{C}|^{2}$,
$\rho^{(2)}_{32}=\rho^{(2)\ast}_{23}\approx-|\Omega_{P}|^{2}\Omega_{C}/2|\Omega_{C}|^{2}\Delta_{1}$,
and
$\rho^{(2)}_{41}=\rho^{(2)\ast}_{14}\approx-\Omega^{\ast}_{C}\Omega_{P}\Omega_{S}/2|\Omega_{C}|^{2}\Delta_{2}$.
And then obtain that:
\begin{eqnarray}
\rho^{(3)}_{31}\approx{|\Omega_{S}|^{2}\over2\Delta_{2}|\Omega_{C}|^{2}}\Omega_{P}-{|\Omega_{P}|^{2}\over2\Delta_{1}|\Omega_{C}|^{2}}\Omega_{P},
\end{eqnarray}
\begin{eqnarray}
\rho^{(3)}_{42}={|\Omega_{P}|^{2}\over4\tilde{\omega}_{42}\tilde{\omega}_{43}-|\Omega_{C}|^{2}}[{1\over\Delta_{1}}-{1\over\Delta_{2}}]\Omega_{S},
\end{eqnarray}
where $\tilde{\omega}_{42}=\Delta_{2}-i\gamma_{42}$ and
$\tilde{\omega}_{43}=\Delta_{2}-\Delta_{1}-i\gamma_{43}$. The
paraxial steady-state propagation equations of $E_{P}$ and $E_{S}$
in the slowly varying envelopes read as:
\begin{eqnarray}
2ik_{P}{\partial\over\partial z}E_{P}=-({\partial^{2}\over\partial
x^{2}}+{\partial^{2}\over\partial
y^{2}})E_{P}-{k^{2}_{P}\over\epsilon_{0}}2N\wp_{31}\rho^{(3)}_{31},
\end{eqnarray}
\begin{eqnarray}
2ik_{S}{\partial\over\partial z}E_{S}=-({\partial^{2}\over\partial
x^{2}}+{\partial^{2}\over\partial
y^{2}})E_{S}-{k^{2}_{S}\over\epsilon_{0}}2N\wp_{42}\rho^{(3)}_{42},
\end{eqnarray}
where $N$ is the density of the atoms. According to Eq. (7), if we
let $|\Delta_{1}|=|\Delta_{2}|$, the propagation dynamics of $E_{S}$
can be treat as linear by the condition of $\Delta_{1},
\Delta_{2}\gg\gamma_{i}, \Omega_{C}$. The reason is: (1)when
$\Delta_{1}=\Delta_{2}$, it is obvious obtained that
$\rho^{(3)}_{42}=0$; (2) when $\Delta_{1}=-\Delta_{2}$,
$\rho^{(3)}_{42}/\rho^{(3)}_{31}\sim|\Omega_{C}|^{2}/\Delta^{2}_{1}\ll1$,
which results in that the refractive index modulation from $E_{P}$
can be neglected. Under this circumstances, $E_{P}$ propagates
nonlinearly with the modulation from $E_{S}$, while $E_{S}$
propagates nearly linearly. This relationship is the key element to
form a discrete system via optical induction \cite{Lederer}.
Therefore, we can apply $|E_{S}|^{2}$ to form the pattern of the
lattice waveguide arrays (See in FIG. 1 (b)). Substituting
$\rho^{(3)}_{31}$ in Eq. (8), we have:
\begin{eqnarray}
i{\partial\over\partial\zeta}U=-{1\over2}\nabla^{2}_{\bot}U
+V(\xi,\eta)U+\kappa|U|^{2}U,
\end{eqnarray}
where
$\nabla^{2}_{\bot}=({\partial^{2}\over\partial\xi^{2}}+{\partial^{2}\over\partial\eta^{2}})$
with $\zeta=zk_{0}$, $\xi=xk_{0}$, $\eta=yk_{0}$;
$U=\Omega_{P}/\Omega_{C}$ is the dimensionless field, and,
\begin{eqnarray}
V(\xi,\eta)=-{N|\wp_{31}|^{2}\over{2\epsilon_{0}\hbar\Delta_{2}}}{|\Omega_{S}(\xi,\eta)|^{2}\over|\Omega_{C}|^{2}},
\kappa={N|\wp_{31}|^{2}\over2\epsilon_{0}\hbar\Delta_{1}}.
\end{eqnarray}
And then, $E_{P}$ experiences two kinds of refractive index
modifications: (1) a periodical index changes induced by
$\Omega_{S}$ (via XPM); (2) an index changes induced by $E_{P}$
itself (via SPM). Here, $\Omega_{S}$ could be viewed as a linear
potential added on $E_{P}$. The sign of detuning $\Delta_{1}$
defines that $E_{P}$ experiencing self-focusing ($\Delta_{1}<0$) or
self-defocusing $\Delta_{1}>0$; while the sign of detuning
$\Delta_{2}$ defines that the energy of the soliton field are
located on a lattice waveguide site (when $\Delta_{2}>0$) or between
lattice sites (when $\Delta_{2}<0$).

In the 1D case, choose the lattices-forming wave to be
$|\Omega_{S}|^{2}=|\Omega^{0}_{S}|^{2}\sin^{2}(\pi\xi/d)$, and the
periodical potential in Eq. (11) can be written as
$V=V^{2}_{0}\sin^{2}(\pi\xi/d)$, where
$V^{2}_{0}=N|\wp_{31}|^{2}|\Omega^{0}_{S}|^{2}/2\epsilon_{0}\hbar\Delta_{2}|\Omega_{C}|^{2}$.
Assuming the probe experiences self-focusing ($\Delta_{1}<0$), the
bound state solution of Eq. (10) can be obtained by the
 imaginary-time propagation (ITP) methods. The parameters are chosen referring to the
Rb atoms as: $\wp_{31}=3.0\times10^{-29}$ D, $\gamma\approx6$ MHz
and $\lambda_{P}=795$ nm. For convenience, other parameters are
chosen as: $d=10$, $|\Omega^{0}_{S}|^{2}/|\Omega_{C}|^{2}=0.1$ with
$\Omega_{C}=2\gamma$, $\Delta_{1}=\Delta_{2}=-100\gamma$ and
$N=3\times10^{15}$ cm$^{-3}$, which are tunable in practice. Then
the normalized parameters $\sigma$ and $V_{0}$ are obtained:
$\sigma=2.5$ and $V_{0}=0.5$. In the simulations, we chose the
normalized transverse power of the field to be
$P=\int|U|^{2}d\xi=0.023$. The numerical solutions are plotted in
FIG. 2(a).

\begin{figure} 
\centering \subfigure[]{
\label{fig_1_a} 
\includegraphics[scale=0.35]{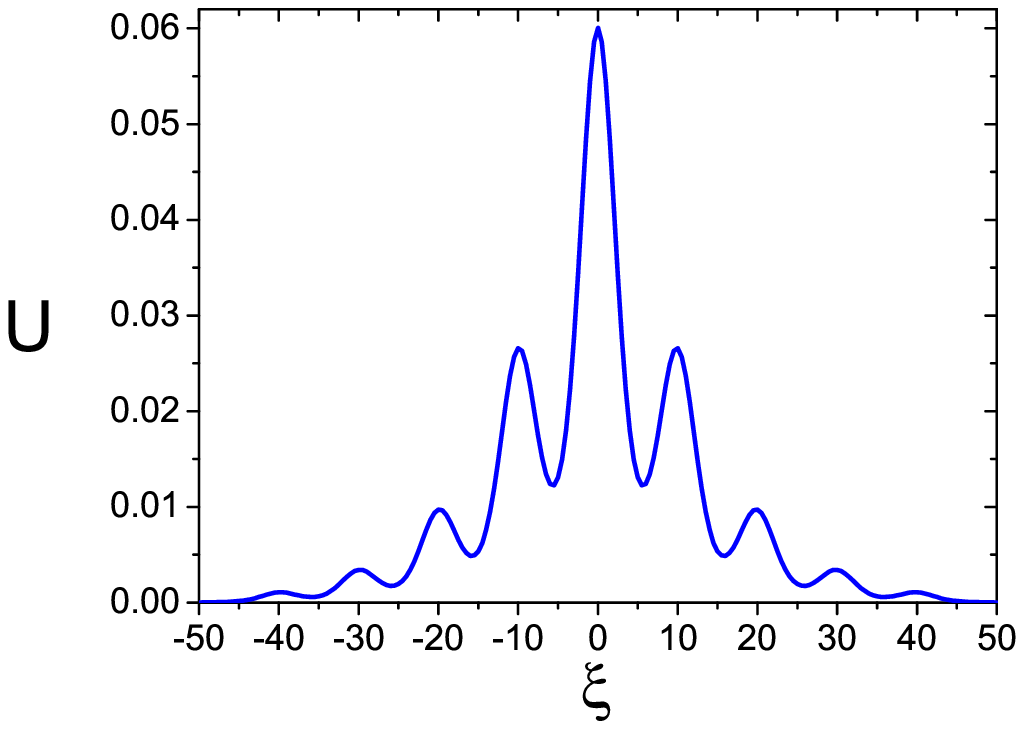}}
\hspace{0.02in} \subfigure[]{
\label{fig_2_b} 
\includegraphics[scale=0.35]{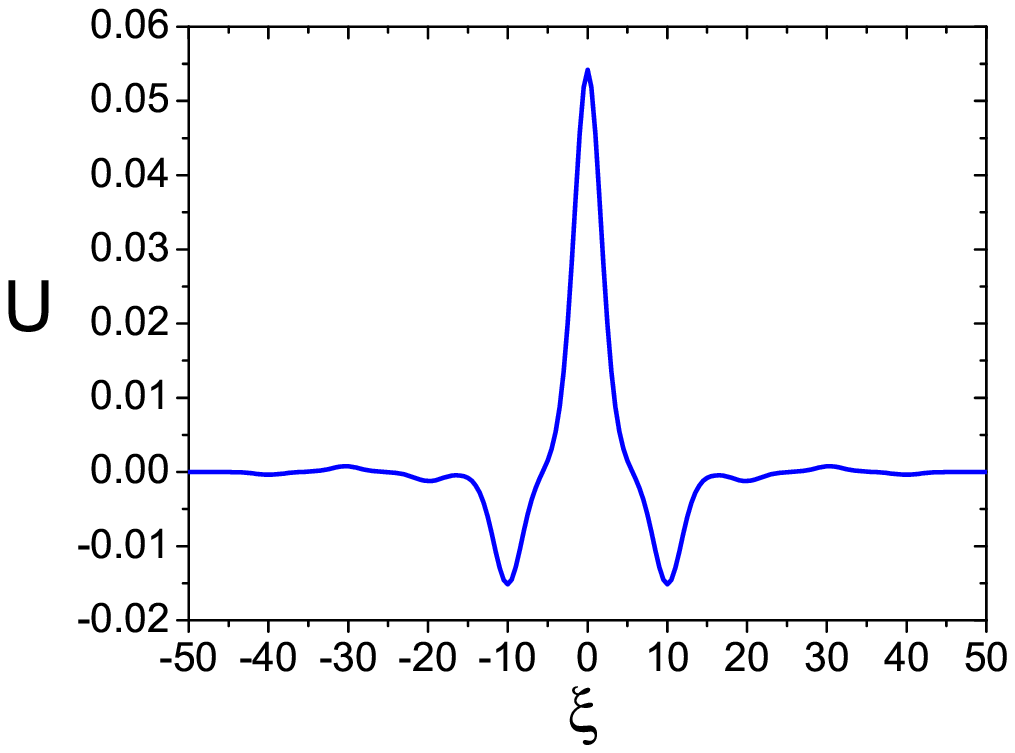}}
\caption{\label{fig:wide}(a)In-phase LS with $\kappa=-2.5$
($\Delta_{1}=-100\gamma$) and $V_{0}=0.5$ and
$P=\int|U|^{2}d\xi=0.023$. (b) $\pi$ out-of-phase LS with
$\kappa=2.5$ ($\Delta_{1}=100\gamma$), $V_{0}=0.75$ and $P=0.0106$.
} \label{fig_2}
\end{figure}

The power density of the soliton can be estimated by
$I_{P}={1\over2}\epsilon_{0}c{\hbar^{2}\Omega^{2}_{C}\over
L_{P}|\wp_{31}|^{2}}P$, where $L_{P}$ is the transverse beam-width
of the probe. Then the photon flux which passes through the soliton
can be estimated by $I_{P}/\hbar\omega_{P}$ \cite{THong}. Moreover,
the group velocity of the probe can also be estimated by \cite{Hau}:
$V_{g}\approx{\hbar
c\epsilon_{0}\over2\omega_{P}}{|\Omega_{C}|^{2}\over|\wp_{31}|^{2}N}$,
 They are all in proportion to $|\Omega_{C}|^{2}$. If we chose
 $\Omega_{C}=2\gamma$, one can obtain that
 $I_{P}=1.81\times10^{-4}$ mW/cm$^{2}$, photon flux is: 7
 mm$^{-2}$ns$^{-1}$, and group velocity is: 3.15 mm/s, which have
 reach the nonclassical light region and the soliton can be termed
 as a NLS.

If the detuning $\Delta_{1}>0$, $E_{P}$ will act as self-defocusing
in the medium. With the periodic modulation induced by $\Omega_{S}$,
Eq. (10) has $\pi$ out-of-phase bright NLS (at the edge of the
Brillouin zone) solutions. Such NLSs can be obtained by angle
incident with the Bloch momentum lies in the vicinity of the edge of
the first Brillouin zone. Diffraction in this condition is negative
and self-defocusing nonlinearity is needed in trapping the fields
\cite{Efremidis,Fleischer2}.  FIG. 2(b) plots a profile of a $\pi$
out-of-phase DS with $\kappa=2.5$ ($\Delta_{1}=100\gamma$),
$V_{0}=0.75$ and $P=0.0106$.

In conclusion, we demonstrate that the optical induction technique
can be applied to build an optical atomic discrete system in an
atomic system through the technique of EIT. The probe field is
experiencing the Kerr nonlinear self modulation and periodical index
modulation of the lattice-forming wave, respectively. By changing
the detuning $\Delta_{1}$, both in-phase and $\pi$ out-of-phase NLSs
can be produced in this system.The technique illustrated in this
paper involves manipulating resonant optical properties of an atomic
medium, and allow people to coherently control the propagation of a
nonclassical light beam. By using this technique, many complex 2D LS
including qusi-crystals LS \cite{Freedman}, defect mediated LS
\cite{LYY1}, honeycomb LS \cite{Peleg}, vortex \cite{Neshev} or
vortex-ring \cite{WXS} LS via optical induction  can all be realized
in such system.

Yongyao Li thanks for Prof. Jianying Zhou for useful discussion.
This work is supported by the NSFC (10934011, 10774193, 10947126,
ZA091201).

\appendix
\section{}
Here, we provide the detail deduction of the results in Eq. (4)-(7).
According to the Eq. (3) and the initial condition:
\begin{eqnarray}
\rho^{(0)}=\left(
\begin{array}{cccc}
 1 & 0 & 0 & 0 \\
 0 & 0 & 0 & 0 \\
 0 & 0 & 0 & 0 \\
 0 & 0 & 0 & 0
\end{array}
\right)
\end{eqnarray}
the slowly varying amplitude evolution equations of the one-step
density matrix elements are read as:
\begin{eqnarray}
\dot{\rho}^{(1)}_{41}=-i\tilde{\omega}_{41}\rho^{(1)}_{41},
\end{eqnarray}
\begin{eqnarray}
\dot{\rho}^{(1)}_{42}=-i\tilde{\omega}_{42}\rho^{(1)}_{42}-{i\over2}\Omega_{C}\rho^{(1)}_{43},
\end{eqnarray}
\begin{eqnarray}
\dot{\rho}^{(1)}_{43}=-i\tilde{\omega}_{43}\rho^{(1)}_{43}-{i\over2}\Omega^{\ast}_{C}\rho^{(1)}_{42},
\end{eqnarray}
\begin{eqnarray}
\dot{\rho}^{(1)}_{44}=-\gamma_{4}\rho^{(1)}_{44},
\end{eqnarray}
\begin{eqnarray}
\dot{\rho}^{(1)}_{31}=-i\tilde{\omega}_{31}\rho^{(1)}_{31}+{i\over2}\Omega_{C}\rho^{(1)}_{21}+{i\over2}\Omega_{P}\rho^{(0)}_{11},
\end{eqnarray}
\begin{eqnarray}
\dot{\rho}^{(1)}_{32}=-i\tilde{\omega}_{32}\rho^{(1)}_{32}-{i\over2}\Omega_{C}(\rho^{(1)}_{33}-\rho^{(1)}_{22}),
\end{eqnarray}
\begin{eqnarray}
\dot{\rho}^{(1)}_{33}=-\gamma_{3}\rho^{(1)}_{33}+{i\over2}(\Omega_{C}\rho^{(1)}_{23}-\Omega^{\ast}_{C}\rho^{(1)}_{32}),
\end{eqnarray}
\begin{eqnarray}
\dot{\rho}^{(1)}_{21}=-i\tilde{\omega}_{21}\rho^{(1)}_{21}+{i\over2}\Omega^{\ast}_{C}\rho^{(1)}_{31},
\end{eqnarray}
\begin{eqnarray}
\dot{\rho}^{(1)}_{22}=-\gamma_{2}\rho^{(1)}_{22}+{i\over2}(\Omega^{\ast}_{C}\rho^{(1)}_{32}-\Omega_{C}\rho^{(1)}_{23}),
\end{eqnarray}
\begin{eqnarray}
\dot{\rho}^{(1)}_{11}=-\gamma_{1}\rho^{(1)}_{11},
\end{eqnarray}
\begin{eqnarray}
\rho^{(1)}_{ij}=\rho^{\ast(1)}_{ji},
\end{eqnarray}
where: $\tilde{\omega}_{41}=\delta+\Delta_{2}-i\gamma_{41}$,
$\tilde{\omega}_{31}=\Delta_{1}-i\gamma_{31}$,
$\tilde{\omega}_{32}=\Delta_{C}-i\gamma_{32}$,
$\tilde{\omega}_{21}=\delta-i\gamma_{21}$, and
$\gamma_{ij}=(\gamma_{i}+\gamma_{j})/2$. Here, we assume
$\gamma_{1}\approx\gamma_{2}\approx0$ and
$\gamma_{3}\approx\gamma_{4}=\gamma$. Because of
$\tilde{\omega}_{21}\approx0$, from Eq. (A6) and (A9), the steady
state solution of $\rho^{(1)}_{31}$ and $\rho^{(1)}_{21}$ are:
\begin{eqnarray}
&&\rho^{(1)}_{31}={2\tilde{\omega}_{21}\Omega_{P}\over4\tilde{\omega}_{31}\tilde{\omega}_{21}-|\Omega_{C}|^{2}}=\rho^{\ast(1)}_{13}\approx0\nonumber\\
&&\rho^{(1)}_{21}={\Omega^{\ast}_{C}\Omega_{P}\over4\tilde{\omega}_{31}\tilde{\omega}_{21}-|\Omega_{C}|^{2}}=\rho^{\ast(1)}_{12}\approx-{\Omega^{\ast}_{C}\Omega_{P}\over|\Omega_{C}|^{2}}.
\end{eqnarray}
And from this set of equations, one can easily obtain other steady
state solution of one-step density matrix elements are all equals 0.
Then the results in Eq. (4) can be obtained.

The slowly varying amplitude evolution equations of the 2-step
density matrix elements are as below:
\begin{eqnarray}
\dot{\rho}^{(2)}_{41}=-i\tilde{\omega}_{41}\rho^{(2)}_{41}+{i\over2}\Omega_{S}\rho^{(1)}_{21},
\end{eqnarray}
\begin{eqnarray}
\dot{\rho}^{(2)}_{42}=-i\tilde{\omega}_{42}\rho^{(2)}_{42}-{i\over2}\Omega_{C}\rho^{(2)}_{43},
\end{eqnarray}
\begin{eqnarray}
\dot{\rho}^{(2)}_{43}=-i\tilde{\omega}_{43}\rho^{(2)}_{43}-{i\over2}\Omega^{\ast}_{C}\rho^{(2)}_{42},
\end{eqnarray}
\begin{eqnarray}
\dot{\rho}^{(2)}_{44}=-\gamma_{4}\rho^{(2)}_{44},
\end{eqnarray}
\begin{eqnarray}
\dot{\rho}^{(2)}_{31}=-i\tilde{\omega}_{31}\rho^{(2)}_{31}+{i\over2}\Omega_{C}\rho^{(2)}_{21},
\end{eqnarray}
\begin{eqnarray}
\dot{\rho}^{(2)}_{32}=-i\tilde{\omega}_{32}\rho^{(2)}_{32}-{i\over2}\Omega_{C}(\rho^{(2)}_{33}-\rho^{(2)}_{22})+{i\over2}\rho^{(1)}_{12},
\end{eqnarray}
\begin{eqnarray}
\dot{\rho}^{(2)}_{33}=-\gamma_{3}\rho^{(2)}_{33}+{i\over2}(\Omega_{C}\rho^{(2)}_{23}-\Omega^{\ast}_{C}\rho^{(2)}_{32}),
\end{eqnarray}
\begin{eqnarray}
\dot{\rho}^{(2)}_{21}=-i\tilde{\omega}_{21}\rho^{(2)}_{21}+{i\over2}\Omega^{\ast}_{C}\rho^{(2)}_{31},
\end{eqnarray}
\begin{eqnarray}
\dot{\rho}^{(1)}_{22}=-\gamma_{2}\rho^{(2)}_{22}+{i\over2}(\Omega^{\ast}_{C}\rho^{(2)}_{32}-\Omega_{C}\rho^{(2)}_{23}),
\end{eqnarray}
\begin{eqnarray}
\dot{\rho}^{(1)}_{11}=-\gamma_{2}\rho^{(2)}_{11},
\end{eqnarray}
\begin{eqnarray}
\rho^{(2)}_{ij}=\rho^{\ast(2)}_{ji}.
\end{eqnarray}
From Eq. (A20) and Eq. (A22), the level $|3\rangle$ and $|2\rangle$
with $\Omega_{C}$ can be treated as a two-level system. Then we
have:
\begin{eqnarray}
\dot{W}^{(2)}_{32}=\dot{\rho}^{(2)}_{33}-\dot{\rho}^{(2)}_{22}=-\Gamma_{32}(W^{(2)}_{32}-W^{(0)}_{32})+i(\Omega_{C}\rho^{(2)}_{23}-\Omega^{\ast}_{C}\rho^{(2)}_{32}),
\end{eqnarray}
where $\Gamma_{32}\approx\gamma_{3}$ is the transverse relaxation
rate between $|2\rangle$ and $|3\rangle$ and
$W^{(0)}_{32}=\rho^{(0)}_{33}-\rho^{(0)}_{22}=0$. Therefore, the
steady state solution of $W^{(2)}_{32}$ can be obtained by:
\begin{eqnarray}
W^{(2)}_{32}=\rho^{(2)}_{33}-\rho^{(2)}_{11}={i\over\Gamma_{32}}(\Omega_{C}\rho^{(2)}_{23}-\Omega^{\ast}_{C}\rho^{(2)}_{32}).
\end{eqnarray}
Substituted Eq. (A26) into Eq. (A19) and its conjugate, under the
steady state condition, one can obtain an equations set about
$\rho^{(2)}_{32}$ and $\rho^{(2)}_{23}$ as below:
\begin{eqnarray}
\begin{cases}
\tilde{\omega}_{32}\rho^{(2)}_{32}-{1\over2}\Omega_{C}W^{(2)}_{32}={1\over2}\Omega_{P}\rho^{(1)}_{12}\\
\tilde{\omega}^{\ast}_{32}\rho^{(2)}_{23}-{1\over2}\Omega^{\ast}_{C}W^{(2)}_{32}={1\over2}\Omega^{\ast}_{P}\rho^{(1)}_{21}
\end{cases}
\end{eqnarray}
The steady state solution of $\rho^{(2)}_{32}$ can be obtained from
this equations set:
\begin{eqnarray}
\rho^{(2)}_{32}=\rho^{\ast(2)}_{23}=-{|\Omega_{P}|^{2}\Omega_{C}\tilde{\omega}_{32}\over2(|\tilde{\omega}_{32}|^{2}+|\Omega_{C}|^{2})|\Omega_{C}|^{2}}
\end{eqnarray}
From Eq. (A14) the steady state solution of $\rho^{(2)}_{41}$ can be
obtained by:
\begin{eqnarray}
\rho^{(2)}_{41}=\rho^{\ast(2)}_{14}=-{\Omega^{\ast}_{C}\Omega_{P}\Omega_{S}\over2\tilde{\omega}_{41}|\Omega_{C}|^{2}}.
\end{eqnarray}

Under the condition of
$\Delta_{1},\Delta_{2}\gg\gamma_{32},|\Omega_{C}|^{2}$,
$\rho^{(2)}_{32}$ and $\rho^{(2)}_{41}$ can be simplified to:
\begin{eqnarray}
\rho^{(2)}_{32}\approx-{|\Omega_{P}|^{2}\Omega_{C}\over2|\Omega_{C}|^{2}\Delta_{1}};
\rho^{(2)}_{41}\approx-{\Omega^{\ast}_{C}\Omega_{P}\Omega_{S}\over2|\Omega_{C}|^{2}\Delta_{2}}.
\end{eqnarray}
Other steady solutions of 2-step density matrix element are all
equal to 0. Then the result of Eq. (5) can be obtained.

Because we only needs the steady solution of $\rho^{(3)}_{31}$ and
$\rho^{(3)}_{42}$ in the 3-step density matrix elements. Therefore
in the 3-step equations, we only need 2 equations sets:
\begin{eqnarray}
\begin{cases}
\dot{\rho}^{(3)}_{42}=-i\tilde{\omega}_{42}\rho^{(3)}_{42}-{i\over2}\Omega_{C}\rho^{(3)}_{43}\\
\dot{\rho}^{(3)}_{42}=-i\tilde{\omega}_{43}\rho^{(3)}_{43}-{i\over2}(\Omega^{\ast}_{C}\rho^{(3)}_{42}+\Omega^{\ast}_{P}\rho^{(2)}_{41}-\Omega_{S}\rho^{(2)}_{23});
\end{cases}
\end{eqnarray}
and
\begin{eqnarray}
\begin{cases}
\dot{\rho}^{(3)}_{31}=-i\tilde{\omega}_{31}\rho^{(3)}_{31}-{i\over2}\Omega_{C}\rho^{(3)}_{21}\\
\dot{\rho}^{(3)}_{21}=-i\tilde{\omega}_{21}\rho^{(3)}_{21}+{i\over2}(\Omega^{\ast}_{C}\rho^{(3)}_{31}-\Omega_{P}\rho^{(2)}_{23}+\Omega^{\ast}_{S}\rho^{(2)}_{41}).
\end{cases}
\end{eqnarray}
By solving these two equations sets, one can obtain the results in
Eq. (6) and Eq. (7).

%

\bibliography{apssamp}

\end{document}